\documentclass[preprint,12pt]{elsarticle}

\usepackage{graphicx}
\usepackage{latexsym}
\usepackage{amsfonts}
\usepackage{amssymb}
\usepackage{amsmath}
\usepackage{bm}
\usepackage{mathrsfs}
\usepackage{slashed}
\usepackage{ascmac}
\usepackage{comment}

\usepackage{dcolumn}

\allowdisplaybreaks[3] 

\usepackage[utf8]{inputenc} 

\newcommand{\cout}[1]{ \if 0 {#1} \fi }
\newcommand{\nn}{\nonumber}
\renewcommand{\=}{&=&}
\newcommand{\nnb}{\nonumber \\}
\newcommand{\pd}{\partial}

\renewcommand{\r}{\right}

\renewcommand{\a}{\alpha}
\renewcommand{\b}{\beta}

\newcommand{\m}{\mu}
\newcommand{\n}{\nu}
\renewcommand{\r}{\rho}
\newcommand{\s}{\sigma}

\newcommand{\Gam}{ \Gamma }
\newcommand{\gam}{ \gamma }

\newcommand{\bB}{{\bm B}}

\newcommand{\bu}{{\bm u}}

\newcommand{\para}{ \parallel}

\newcommand{\zero}{{(0)}}
\newcommand{\one}{{(1)}}

\usepackage{color} 
\usepackage[normalem]{ulem}  

\begin{document}

\begin{frontmatter}



\title{First-order spin magnetohydrodynamics}

\author{Zhe Fang}
\ead{12345071@zju.edu.cn}
\affiliation{Zhejiang Institute of Modern Physics, Department of Physics, Zhejiang University, Hangzhou, Zhejiang 310027, China}

\author{Koichi Hattori}
\ead{koichi.hattori@zju.edu.cn}
\affiliation{Zhejiang Institute of Modern Physics, Department of Physics, Zhejiang University, Hangzhou, Zhejiang 310027, China}
\affiliation{Research Center for Nuclear Physics, Osaka University, 
10-1 Mihogaoka, Ibaraki, Osaka 567-0047, Japan}

\author{Jin Hu}
\ead{hu-j23@fzu.edu.cn}
\affiliation{Department of Physics, Fuzhou University, Fujian 350116, China}


\begin{abstract}
Based on recent papers, we discuss formulation of the first-order relativistic spin magnetohydrodynamics (MHD) with the totally antisymmetric spin current and properties of the anisotropic linear waves awaken near an equilibrium configuration. 
We show that there appears a critical angle in the momentum direction of the linear waves, where a pair of propagating modes turns into purely diffusive modes. 
Due to this critical behavior, polynomial solutions do not fully capture the angle dependence of the linear waves. 
\end{abstract}
%

\begin{keyword} Hydrodynamics \sep Strong magnetic field 
\sep Spin transport \sep Anisotropic systems



\end{keyword}

\end{frontmatter}



\section{Introduction}\label{s1}

We discuss relativistic spin magnetohydrodynamics (MHD) with dissipative effects from the first-order derivative expansion based on the recent papers \cite{Fang:2024skm,Fang:2024sym}. 
This framework describes the interplay between fluid dynamics and a dynamical magnetic field, giving rise to the Alfven and magneto-sonic waves. 
We further include spin current in the framework. 
We obtain the complete set of dispersion relations for the anisotropic linear waves with the dissipative effects.

Strong magnetic fields induce intriguing transport phenomena 
such as the chiral magnetic effect discussed across research fields such as relativistic heavy-ion collisions, condensed matter physics, and cosmology/astrophysics (see, e.g., Refs.~\cite{Kharzeev:2015znc, Hattori:2016emy,Yan:2016euz,Armitage:2017cjs,Kamada:2022nyt} for reviews and references therein). 
One can describe these phenomena with the low-energy effective theory, i.e., (chiral) hydrodynamics developed in recent years \cite{Son:2009tf, Grozdanov:2016tdf,Hattori:2017usa,Hongo:2021ona,Matsumoto:2022lyb, Brandenburg:2023aco} (see Ref.~\cite{Hattori:2022hyo,Hattori:2023egw} for reviews).


Dynamics of angular momentum is another hot topic triggered by recent measurements for spin polarization/alignment of hadron species emitted from quark-gluon plasma (QGP)  
\cite{STAR:2017ckg, STAR:2018gyt, mohanty_spin_2021,STAR:2019erd,STAR:2020xbm, ALICE:2022dyy,Micheletti:2023qlh}. 
A nonzero angular momentum, which is mechanically created 
by a non-central collision of two large nuclei, is converted to 
thermal vorticity in QGP and spin of hadrons in the final state \cite{Liang:2004ph,Liang:2004xn,Betz:2007kg,Becattini:2007sr,Gao:2007bc,Becattini:2016gvu}. 
The experimental results on the local $\Lambda$  spin polarization pose a spin sign puzzle that cannot be solely explained by thermal effects \cite{Becattini:2021iol, Wu:2019eyi,Becattini:2024uha}. 
This deviation may be attributed to the non-equilibrium evolution of spin degrees of freedom, inspiring theoretical developments in spin hydrodynamics \cite{Hattori:2019lfp,Fukushima:2020ucl,Gallegos:2021bzp,Li:2020eon,Hu:2021lnx,Hu:2022azy,Singh:2022ltu,Cao:2022aku,Daher:2022wzf,Sarwar:2022yzs,Kiamari:2023fbe,Xie:2023gbo,Ren:2024pur}.

Recently, magnetic-field effects were discussed as sources of not only the spin splitting between hadrons and anti-hadrons \cite{Xu:2022hql, Peng:2022cya,Buzzegoli:2022qrr} but also the local spin polarization \cite{Sun:2024isb}. 
The interplay between the magnetic fields and spin dynamics in QGP draws considerable interest.

In this article, we closely discuss the dispersion relations of 
the anisotropic linear waves in a magnetic field. 
We emphasize the existence of a critical angle where a pair of propagating modes turns into purely diffusive modes. 
This critical behavior is already seen in MHD without spin \cite{Fang:2024skm} 
and is not captured by polynomial solutions. 
Therefore, the solutions in the small-momentum expansion break down 
as the momentum direction approaches the critical angle. 
We investigate the behavior of the diffusive regime 
and provide an alternative form of solution.


We use the mostly plus metric convention ${\eta ^{\mu \nu }} = \text{diag}( - 1,1,1,1)$ and the completely antisymmetric tensor with the convention $\epsilon^{0123} =  + 1$. 
Then, the fluid velocity ${u^\mu }$ is normalized 
as ${u^\mu }{u_\mu } = - 1$. 
We define the projection operator ${\Delta ^{\mu \nu }} = {\eta ^{\mu \nu }} + {u^\mu }{u^\nu }$ such that $u_\mu \Delta^{\mu\nu}=0 $.

\section{Formulation of spin magnetohydrodynamics}\label{s2}

Hydrodynamics describes the low-energy dynamics, 
where the relevant degrees of freedom are gapless modes. 
These modes are associated with 
the conserved charges surviving in a long spacetime scale. 
Therefore, equations of motion are given by a set of conservation laws 
informed by symmetries of a system.

The conservation laws deduced from the translational, 
rotational, and one-form symmetries read \cite{Fang:2024skm,Fang:2024sym} 
\begin{equation}
{\partial _\mu }{\Theta ^{\mu \nu }} = 0 , \quad 
{\partial _\mu }{J^{\mu \a\b }} = 0 , \quad 
{\partial _\mu }{{\tilde F}^{\mu \nu }} = 0 , 
\label{con4}
\end{equation}
where $ \Theta ^{\mu \nu }$, $ J^{\mu \a\b }$, ${\tilde F}^{\mu \nu } $ are the energy-momentum tensor, 
the angular momentum tensor, and 
the dual electromagnetic field strength tensor, respectively. 
The last one is the magnetic-flux conservation law. 
The angular momentum tensor $J^{\mu\a\b} $ can be decomposed 
into the contributions of the spin current $\Sigma^{\mu\a\b} $ 
and the orbital angular momentum as 
\begin{eqnarray}
J^{\mu\a\b} = \Sigma^{\mu\a\b}
+ x^\a \Theta^{\mu\b} - x^\b \Theta^{\mu\a}
\, .
\end{eqnarray}
Then, the total angular momentum conservation in Eq.~(\ref{con4}) is cast into another form 
\begin{eqnarray}
\label{eq:spin-cons}
\partial _\mu \Sigma^{\mu \a\b } =  - 2 \Theta^{[\a\b]}
\, .
\end{eqnarray}
Spin is not a conserved quantity by itself in relativistic systems. 
The antisymmetric part of the energy-momentum tensor 
quantifies the torque exerted on a fluid cell that gives rise to conversion between the spin and orbital components of the angular momentum. 
Since the spin density is not a strictly conserved quantity, 
spin hydrodynamics should be understood as an extended 
hydrodynamic framework with a quasihydrodynamic mode \cite{Hattori:2019lfp}. 

The local conservation laws of the energy-momentum tensor 
and the angular momentum tensor are preserved 
under the simultaneous shifts of ${\Theta ^{\mu \nu }} $ 
and $J^{\mu\a\b} $ transformations. 
Then, the separation between the spin and orbital angular momentum is not unique. 
In the end, the definitions of the spin and orbital angular momentum may obey specific measurement processes. 
Here, we assume that $\Sigma^{\mu\alpha\beta}$ is 
a totally antisymmetric tensor.


The temporal components of these currents 
provide the density of corresponding conserved charges 
\begin{eqnarray}
\label{eq:charges}
e =  u_\mu u_\nu {\Theta ^{\mu \nu }} , \quad 
S^{\a\b} = - u_\mu \Sigma^{\mu \a\b } , \quad 
B^\mu  = -  {\tilde F}^{\mu \nu } u_\nu . 
\end{eqnarray}
The system of equations (\ref{con4}) is closed with the help of 
the constitutive equations that express the spatial components of 
the conserved currents as functionals of the conserved charges. 
The constitutive equations can be determined so that 
their forms are consistent with the first and second laws of thermodynamics as \cite{Fang:2024sym} 
\begin{subequations}
\label{con6}
\begin{eqnarray}  
{\Theta ^{\mu \nu }} \= e{u^\mu }{u^\nu } + {p }{\Xi ^{\mu \nu }} 
+ \Big( p - \frac{B^2}{\mu_m} \Big) {b^\mu }{b^\nu }
\nnb
&&- T  \left(
\begin{array}{ll}
\eta^{\m\n\r\s} & \xi^{\m\n\r\s}
\\
\xi^{\prime \m\n\r\s} & \gam^{\m\n\r\s}
\end{array}
\right)
 \left(
\begin{array}{l}
\pd_{(\r} \beta_{\s)}  
\\
\pd_{[\r} \beta_{\s]} - 2\beta  \omega_{\r\s}
\end{array}
\right) , \\
{\Sigma ^{\mu \alpha \beta }} \= u^\mu S^{\alpha\beta} - u^\alpha S^{\mu\beta}+u^\beta S^{\mu\alpha} 
,\\
{{\tilde F}^{\mu \nu }} \= {B^\mu }{u^\nu }  - {B^\nu }{u^\mu } 
- T{\rho ^{\mu \nu \rho \sigma }}{\partial_ {[\rho }}(\beta {H_{\sigma ]}}) ,
\end{eqnarray}
\end{subequations} 
where we introduced a unit vector ${b^\mu } = {B^\mu }/\sqrt {{B^\nu }{B_\nu }}$ and the projection operator ${\Xi ^{\mu \nu }} = {\Delta ^{\mu \nu }} - {b^\mu }{b^\nu }$ 
such that $b_\mu \Xi ^{\mu \nu }=0= u_\mu \Xi ^{\mu \nu }$.  
We assume the linear relations between the conserved charges 
and the Lagrange multipliers, i.e.,  
$H^\mu= B^\mu/\mu_m $ and $\omega^{\mu\nu}=S^{\mu\nu}/\chi$ 
with the magnetic permeability $ \mu_m$ 
and the spin susceptibility $\chi $ being constants in spacetime. 
We choose to work with the Landau frame, so that 
the heat current is vanishing $h^{\mu} =0 $. 
Also, the rotational heat current introduced in Ref.~\cite{Hattori:2019lfp} is higher order in derivative in the totally antisymmetric pseudogauge (see Ref.~\cite{Fang:2024sym}). 
Onsager's reciprocal relation states the relations among the viscous tensors 
$ \eta^{\m\n\r\s}(b^\mu) = \eta^{\r\s\m\n} (-b^\mu) $, 
$ \gam^{\m\n\r\s}(b^\mu) = \gam^{\prime\r\s\m\n} (-b^\mu) $, 
$ \xi^{\m\n\r\s}(b^\mu) = \xi^{\r\s\m\n} (-b^\mu) $, 
and $ \xi^{\m\n\r\s}(b^\mu) = \xi^{\prime\r\s\m\n} (-b^\mu) $.

One can use the available tensor $b^\mu b^\nu$ and $ \Xi^{\mu\nu}$ 
to construct the viscous tensors $ {\eta ^{\mu \nu \rho \sigma }} $, 
$ {\gam ^{\mu \nu \rho \sigma }} $, and $ {\xi ^{\mu \nu \rho \sigma }} $ 
and the resistivity tensor $ {\rho ^{\mu \nu \rho \sigma }} $ . 
The general forms allowed by the thermodynamic stability 
are found to be \cite{Fang:2024sym} 
\begin{subequations}
\label{con16}
\begin{eqnarray}  
{\eta ^{\mu \nu \rho \sigma }}  
\= 
\begin{pmatrix}
b^\mu b^\nu & \Xi^{\mu\nu}
\end{pmatrix}
\begin{pmatrix}
\zeta_\para & \zeta_\times \\
\zeta'_\times & \zeta_\perp
\end{pmatrix}
\begin{pmatrix}
b^\rho b^\sigma \\ \Xi^{\rho\sigma}
\end{pmatrix} 
\\
&&
+ 2{\eta _\parallel }({b^\mu }{\Xi ^{\nu (\rho }}{b^{\sigma )}} + {b^\nu }{\Xi ^{\mu (\rho }}{b^{\sigma )}}) 
 + {\eta _ \bot }({\Xi ^{\mu \rho }}{\Xi ^{\nu \sigma }} + {\Xi ^{\mu \sigma }}{\Xi ^{\nu \rho }} - {\Xi ^{\mu \nu }}{\Xi ^{\rho \sigma }}) 
 ,
\nnb 
{\gamma ^{\mu \nu \rho \sigma }} 
\= {\gamma _ \bot }({\Xi ^{\mu \rho }}{\Xi ^{\nu \sigma }} - {\Xi ^{\mu \sigma }}{\Xi ^{\nu \rho }})  
- 2{\gamma _\parallel }({b^\mu }{\Xi ^{\nu [\rho }}{b^{\sigma ]}} - {b^\nu }{\Xi ^{\mu [\rho }}{b^{\sigma ]}})
,
\\
{\xi ^{\mu \nu \rho \sigma }} 
\= 2{\xi _\parallel }({b^\mu }{\Xi ^{\nu [\rho }}{b^{\sigma ]}} + {b^\nu }{\Xi ^{\mu [\rho }}{b^{\sigma ]}}  
+ {b^\mu }{\Xi ^{\nu (\rho }}{b^{\sigma )}} - {b^\nu }{\Xi ^{\mu (\rho }}{b^{\sigma )}}), 
\\
    {\rho ^{\mu \nu \rho \sigma }} \= - 2{\rho_ \bot }({b^\mu }{\Xi ^{\nu [\rho }}{b^{\sigma ]}} - {b^\nu }{\Xi ^{\mu [\rho }}{b^{\sigma ]}})  +  2{\rho_\parallel }{\Xi ^{\mu [\rho }}{\Xi ^{\sigma ]\nu }}, 
    \label{add1}
\end{eqnarray}
\end{subequations}
where $\zeta_\times = \zeta_\times' $ by virtue of Onsager's reciprocal relation. 
The semi-positivity of the entropy current requires the inequalities 
\begin{eqnarray}
\label{inequality}
&&  {\zeta _ \bot } \ge 0, \quad 
{\zeta _\parallel } \ge 0, \quad 
{\zeta _\parallel }{\zeta _ \bot } \ge \zeta _ \times ^2, \quad 
 {\eta _\parallel } \ge 0,  \quad 
{\eta _ \bot } \ge 0, \quad 
\nnb
 && 
 {\gamma _\parallel } \ge 0, \quad 
{\gamma _ \bot } \ge 0, \quad 
{\gamma _\parallel }{\eta _\parallel } \ge {\xi _\parallel^2 }
, \quad 
\label{con20} 
\\
&&
\rho_\perp \ge 0, \quad \rho_\parallel\ge 0
. \nn
\end{eqnarray}

\begin{figure}[t]
\begin{center}
   \includegraphics[width=\hsize]{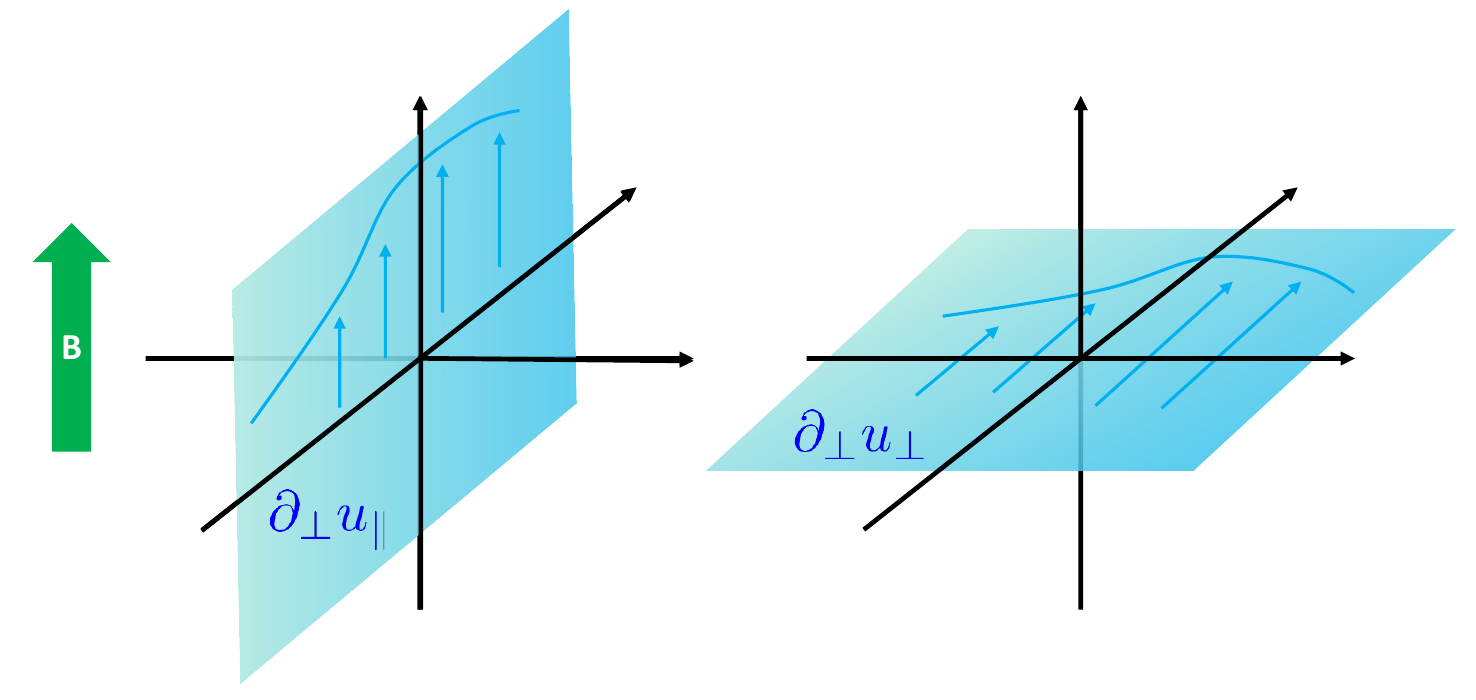}
\end{center}
\vspace{-1cm}
\caption{Two components of the shear viscosity in a magnetic field. 
}
  \label{fig:shear}
\begin{minipage}{0.45\hsize} 
	\begin{center} 
\includegraphics[width=\hsize]{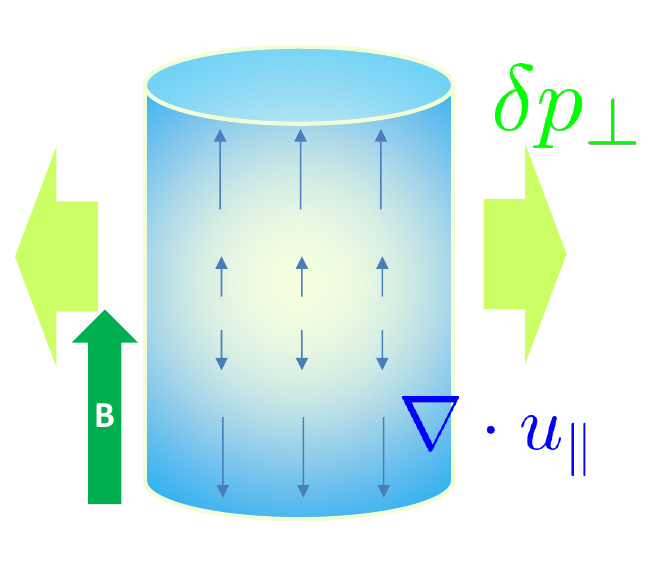}
	\end{center}
\vspace{-0.5cm}
\caption{Cross bulk viscosity in a magnetic field}
\label{fig:bulk}
\end{minipage}
\begin{minipage}{0.45\hsize}
	\begin{center}
\includegraphics[width=\hsize]{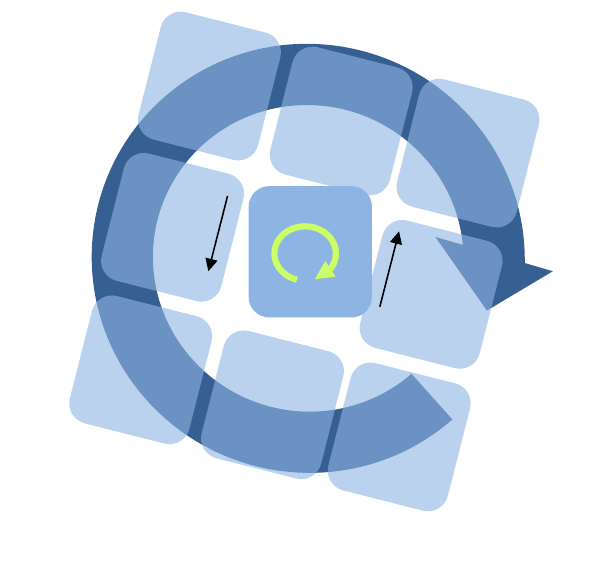}
	\end{center}
\vspace{-1cm}
\caption{Rotational viscosity from the ``non-slip'' condition}
\label{fig:rotation}
\end{minipage}
\end{figure}

The resistivity tensor $ {\rho ^{\mu \nu \rho \sigma }} $ provides the constitutive equations of an induced electric field and current that have two components in parallel and perpendicular to the magnetic field (see Ref.~\cite{Hattori:2022hyo} for a review). 
There is no Hall term in neutral systems. 
The physical meaning of the viscous tensors is as follows. 
The symmetric viscous tensor $ {\eta ^{\mu \nu \rho \sigma }} $ 
contains the shear and bulk components. 
The shear viscosity is split into two components $ \eta_{\para,\perp}$
in parallel and perpendicular to a magnetic field (see Fig.~\ref{fig:shear}). 
The bulk viscosity is also split into two components $ \zeta_{\para,\perp}$. 
There is one more bulk viscosity $ \zeta_\times$ that quantifies 
the off-equilibrium pressure in the parallel (perpendicular) direction 
in response to a fluid expansion in the perpendicular (parallel) direction 
(see Fig.~\ref{fig:bulk}).

The antisymmetric viscous tensor $ {\gam ^{\mu \nu \rho \sigma }} $ 
provides the rotational viscosity introduced 
in spin hydrodynamics \cite{Hattori:2019lfp}. 
The rotational viscosity quantifies the friction 
between a rotating fluid cell and the surrounding rotational motion 
of fluid (see Fig.~\ref{fig:rotation}). 
The rotational viscosity is also split into two components in a magnetic field. 
Finally, there is a cross term between the symmetric and antisymmetric parts 
of the energy-momentum tensor given by $ {\xi ^{\mu \nu \rho \sigma }} $. 
This viscous term converts the shear to the antisymmetric torque 
and the vorticity to the symmetric stress.

\section{Anisotropic linear waves}

\subsection{Linearized equations near an equilibrium} \label{s3}

In this section, we derive the linearized first-order hydrodynamic equations 
for small perturbations near an equilibrium state specified as 
$u^\mu  = (1,0,0,0)$, $B^\mu  = (0,0,0,B)$, 
and $ S^{\mu \nu } = 0 $.  
The conserved charges are displaced from their equilibrium values as 
\begin{eqnarray}
&& e \to e+ \delta e (x), \quad 
 {u^\mu } \to u^\mu  + \delta {u^\mu }(x) ,
\label{don01}
 \\
 &&
{B^\mu } \to B^\mu  + \delta {B^\mu } (x), \quad 
{S^{\mu \nu }} \to { S^{\mu \nu }} + \delta {S^{\mu \nu }} (x) 
\nn
.
\end{eqnarray}
We will linearize the hydrodynamic equations with respect to these perturbations. 
One can extract the three independent degrees of freedom 
of $S^{\mu\nu}$ by a transformation to a spatial vector 
$ \sigma^\mu=-\frac 12\epsilon^{\mu\nu\rho\sigma} u_\nu S_{\rho\sigma}$.

We introduce a momentum representation of the perturbation  
\begin{eqnarray}
\label{eq:linear-solutions}
     \delta \bu (t,x,z) =  
     \delta \tilde \bu(\omega,k_\perp,k_\para) e^{- i \omega t + i k_\perp x + i k_\para z}
     ,
\end{eqnarray}
and the same for $\delta e $ and $\delta \bB $. 
Here, without loss of generality, 
we have set the transverse coordinate system in such a way that 
the dependence on the $y$ coordinate vanishes. 
Then, the linearized equation can be summarized in matrix equations 
\begin{eqnarray} 
\label{eq:matrix-eqs}
\Big( \, A_\zero + i A_\one \, \Big) 
\begin{pmatrix}
  \delta \tilde B_y  \\ \delta \tilde u_y  \\ \delta {\tilde \sigma _x}
  \\ \delta {\tilde \sigma _z}
\end{pmatrix} 
=0 , \quad 
\Big( \, M_\zero + i M_\one \, \Big) 
\begin{pmatrix}
\delta \tilde \epsilon \\ \delta \tilde u_x \\  
\delta \tilde u_z \\ \delta \tilde B_x \\ \delta \tilde B_z \\
 \delta {\tilde \sigma _y}
\end{pmatrix}
=0 .
\end{eqnarray} 
The first set of matrices is given as \cite{Fang:2024skm, Fang:2024sym}
\begin{subequations} 
\label{eq:matrix-Alfven}
\begin{eqnarray}
A_\zero \= 
\begin{pmatrix}
 \omega   &  B k_\para 
\\
 h \frac{v_A^2}{B} k_\para &  h \omega 
 \\
& & \omega + i \Gam_\para & 
\\
& & & \omega + i \Gam_\perp
\end{pmatrix} 
,  
\\
A_\one \=  
\begin{pmatrix}
  \rho_\perp' k_\para^2 + \rho_\para '  k_\perp^2  &  & &
\\
   & ( \eta_\para +\gam_\para-2\xi_\para )  k_\para^2 
   + ( \eta_\perp+\gam_\perp )  k_\perp^2 & 
 \frac{4i}{\chi}(\gam_\para-\xi_\para)k_\para  &
- \frac{i}{2} \Gam_\perp k_\perp 
  \\
& -2i(\gam_\para - \xi) k_\para &  0& 0
\\
& 2i \gam_\perp k_\perp &   0& 0
\end{pmatrix}
. \nnb
\end{eqnarray} 
\end{subequations} 
The second set is given as 
\begin{subequations}
\label{eq:matrix-sonic}
\begin{eqnarray}  
\hspace{-1.2cm} 
M_\zero \=  
\begin{pmatrix} 
0 & - k_\perp  B & 0  & 0 & \omega    & 0
\\
0 &  k_\para B & 0 &  \omega  & 0    & 0
\\
\omega & -  h k_\perp &  
 h  ( v_A^2  - 1) k_\para & 0 & h \frac{v_A^2}{B}  \omega    & 0
\\
- c_s^2 k_\perp &  h \omega & 0 &
h \frac{v_A^2}{ B} k_\para & - h \frac{v_A^2}{ B}   k_\perp    & 0
\\
- c_s^2 k_\para & 0 &  -  h  ( v_A^2  - 1) \omega & 0 &  0  & 0
\\
0&0&0&0&0& \omega + i \Gam_\para
\end{pmatrix} 
,
\\ \label{eq:M1}
\hspace{-1.2cm}
M_\one \= 
\begin{pmatrix} 
-   \frac{ B c_s^2 }{ h ( 1 -v_A^2) } 
\rho'_\perp  k_\perp^2
 & 0 & 0  & - \rho_\perp'  k_\para k_\perp  
&  \rho_\perp' k_\perp^2   & 0
\\
\frac{ B c_s^2 }{ h ( 1 -v_A^2) } 
\rho'_\perp  k_\perp k_\para 
 & 0 & 0  &  \rho_\perp' k_\para^2 
& - \rho_\perp' k_\perp k_\para    & 0
\\ 
0 & 0& 0 & 0& 0    & 0
\\
0 & m_{11} &m_{12} &  0& 0 &
- i \frac{4}{\chi} (\gam_\para - \xi_\para) k_\para
\\
0 & m_{21} & m_{22} &  0& 0 &
i \frac{4}{\chi} (\gam_\para+\xi_\para) k_\perp 
\\
0 &m_{31} & m_{32} &
0 & 0& 0&
\end{pmatrix}
, \nnb
\end{eqnarray}
\end{subequations}  
where 
\begin{eqnarray}
\begin{pmatrix}
m_{11} & m_{12} \\
m_{21} & m_{22} \\
m_{31} & m_{32}
\end{pmatrix}
=
\begin{pmatrix}
(\zeta_\perp + \eta_\perp  ) k_\perp^2 
+ (\eta_\para + \gam_\para - 2\xi_\para ) k_\para^2 &
(\zeta_\times + \eta_\para -\gam_\para ) k_\perp k_\para
\\
(\zeta_\times + \eta_\para -\gam_\para) k_\perp k_\para &
\zeta_\para k_\para^2 + (\eta_\para +\gam_\para+2\xi_\para ) k_\perp^2 
\\
2i(\gam_\para-\xi_\para) k_\para &-2i(\gam_\para+\xi_\para) k_\perp 
\end{pmatrix}
.
\end{eqnarray}
We assumed that the contributions of 
the matter and magnetic components to the equilibrium 
energy density and pressure can be separated as 
\begin{eqnarray}
 p_\perp=P+\frac{B^2}{2\mu_m}, \quad 
e=\epsilon+\frac{B^2}{2\mu_m} ,
\label{relations}
\end{eqnarray}
where $P$ and $\epsilon$ are the equilibrium energy density and pressure from the matter component. 
Then, the enthalpy is given as 
$   h =   e +   p =  \epsilon +   P +   B^2 /\mu_m $. 
We defined the Alfven velocity $ v_A$ and the sound velocity $ c_s $ as 
\begin{eqnarray}
v_A = \frac{B}{\sqrt{  \mu_m h }} , \quad 
c_s = \sqrt{ \frac{\delta P}{ \delta \epsilon } } .
\end{eqnarray}
Also, we use the shorthand notations 
\begin{eqnarray}
    \rho'_\para = \frac{ \rho_\para}{\mu_m} , \quad 
     \rho'_\perp = \frac{ \rho_\perp}{\mu_m} , \quad 
     \Gam_\para =\frac{8\gamma_\parallel}{\chi} , \quad 
\Gam_\perp =\frac{8\gamma_\perp}{\chi}.
\end{eqnarray}


\cout{

We have obtained the linearized equations for spin MHD, and it is useful to note that the equations obtained above can be divided into two groups. One of them contains the variables $ (\delta {u_y},\delta {B_y},\delta {\sigma _x},\delta {\sigma _z})$, while the other contains $( \delta \epsilon,\delta {u_x},\delta {u_z},\delta {B_x},\delta {\sigma _y})$. 
The former group is summarized as 
\begin{subequations}\label{f2}
    \begin{eqnarray}
0\=h{\partial _0 }{\delta u_y } - 
 \frac{h}{B} v_A^2  {\partial_z }\delta {B_y}
\\
&&
-[(\eta_\perp+\gamma_\perp)\partial_x^2+(\eta_\parallel+\gamma_\parallel-2\xi_\parallel)\partial_z^2]\delta u_y
\nnb
&&
- \frac 12 \Gam_\perp 
\partial_x \delta \sigma_{z}
+\frac{4}{\chi}(\gamma_\parallel-\xi_\parallel)\partial_z\delta\sigma_{x},
\nnb
0\=B\partial_z \delta u_y-\partial_0 \delta B_y+ \rho'_\perp \partial_z^2 \delta B_y+\rho'_\parallel\partial_x^2 \delta B_y,
\\
0\=(\Gam_\para+\partial_0)\delta\sigma_x
- 2(\gamma_\parallel-\xi_\parallel)\partial_z\delta u_y,
\\
0\=(\Gam_\perp+\partial_0)\delta\sigma_z
+ 2\gamma_\perp\partial_x\delta  u_y .
    \end{eqnarray}
\end{subequations}
We assume a single-mode solution 
for the flow perturbation 
\begin{eqnarray}
    \delta \bu (t,x,z) =  \delta \tilde \bu(\omega,k_\perp,k_\para) e^{- i \omega t + i k_\perp x + i k_\para z}
    , \label{eq:mode-exp}
\end{eqnarray}
and similar forms for other perturbations. 
Then, these equations can be summarized in an algebraic matrix form
\begin{eqnarray}
    0=M_1 (\delta {\tilde u_y},\delta {\tilde B_y},\delta {\tilde \sigma _x},\delta {\tilde \sigma _z})^T, \label{quartic eq1}
\end{eqnarray}
where $M_1$ is the coefficient matrix. 
On the other hand, the latter group reads 
\begin{subequations}
\label{f1}
    \begin{eqnarray}
0 \=  \partial_0 \delta \epsilon
+ \frac{h}{B} v_A^2  \partial_0\delta B+h\partial_x \delta u_x
+ h (1-v_A^2) \partial_z \delta u_z,\nnb
\\
0\=h{\partial _0 }{\delta u_x } + c_s^2{\partial _x }{\delta \epsilon } - \frac{h}{B} v_A^2  
( {\partial_z }\delta {B_x}- \partial_x\delta B_z)
\nnb
&&
-[(\zeta_\perp+\eta_\perp)\partial_x^2+(\eta_\parallel+\gamma_\parallel-2\xi_\parallel)\partial_z^2]\delta u_x
\nnb
&&
-(\zeta_\times+\eta_\parallel-\gamma_\parallel)\partial_z\partial_x\delta u_z
- \frac 4\chi(\gamma_\parallel-\xi_\parallel)\partial_z\delta\sigma_{y},
\\
0\=c_s^2{\partial _z }\delta \epsilon
+ h (1-v_A^2){\partial _0 }{\delta u_z } 
-(\zeta_\times+\eta_\parallel-\gamma_\parallel)\partial_z \partial_x\delta u_x
\nnb
&&
-(\zeta_\parallel\partial_z^2
+(\eta_\parallel+\gamma_\parallel+2\xi_\parallel)\partial_x^2)\delta u_z
\nnb
&&
+\frac{4}{\chi}(\xi_\parallel+\gamma_\parallel)\partial_x\delta\sigma_{y},
\\
0\=B\partial_z \delta u_x-\partial_0 \delta B_x
\nnb
&&
- \rho'_\perp[\partial_z\partial_x \delta B_z-\frac{c_s^2 }{1-v_A^2} \frac{B}{h} \partial_z\partial_x \delta \epsilon-\partial_z^2 \delta B_x],
\\
0\=
( \Gam_\para +\partial_0)\delta\sigma_y
\nnb
&&
+ 2[(\gamma_\parallel - \xi_\parallel)\partial_z\delta u_x - (\gamma_\parallel+\xi_\parallel)\partial_x\delta u_z]
.
    \end{eqnarray}
\end{subequations}

}

\subsection{Solutions for the linearized equations}\label{s4}

Searching analytic solutions for the linearized equations become 
challenging as the symmetry is reduced and 
the number of degrees of freedom becomes large. 
Spin MHD is one such case with nine degrees of freedom 
and a broken spatial rotational symmetry.

We developed a general and simple analytic algorithm 
for the solution search in Ref.~\cite{Fang:2024skm} 
and obtained the full analytic solution for the first-order MHD. 
This method works on an order-by-order basis in an arbitrary expansion parameter. 
We then applied this method to spin MHD in Ref.~\cite{Fang:2024sym}. 
The relevant expansion parameter is the momentum $k $ as in usual hydrodynamics. 
However, we find that the small-momentum expansion breaks down in MHD 
when the momentum direction approaches the perpendicular direction to 
an equilibrium magnetic field. 
In this region, a trigonometric function for the angle dependence serves as another expansion parameter as shown below.

In the momentum expansion, we search solutions up to the order $ k^2$ 
since the higher terms are not under control within the first-order hydrodynamics. 
We find the solutions for the first equation in Eq.~(\ref{eq:matrix-eqs}) 
in the following forms 
\begin{subequations}
\begin{eqnarray}
\label{quantic NLO solution1}
\omega \=\pm v_A k \cos\theta-i U_{1} k^2 ,
\\
\label{quantic NLO solution2}
\omega \= -i\Gamma_\parallel -i U_2  k^2 ,
\\
\label{quantic NLO solution3}
\omega \= -i\Gamma_\perp -i U_3 k^2 .
\end{eqnarray}\label{quantic NLO solution}
\end{subequations}
with the momentum decomposition $k_\para=k \cos \theta$ and $k_\perp=k\sin\theta$. 
The corrections at the order $ k^2$ are found to be
\begin{subequations}
    \begin{eqnarray}
    \label{NLO11}
U_1 \= \frac{1}{2h}\big[( \tilde \eta_\para +h\rho_\perp')\cos^2\theta 
+(\eta_\perp+h\rho_\parallel')\sin^2\theta\big], 
    \\
    \label{NLO12}
U_2 \=\frac{(\gamma_\parallel-\xi_\parallel)^2}{h\gamma_\parallel}\cos^2\theta, \,\,
    \\
U_3 \= \frac{\gamma_\perp}{h} \sin^2\theta.
\end{eqnarray}\label{NLO}
\end{subequations}
Here, we defined a specific combination of the viscous coefficients 
\begin{eqnarray}
\label{eq:eta-tilde}
\tilde \eta_\parallel =    \eta_\parallel-\frac{\xi^2_\parallel}{\gamma_\parallel} .
\end{eqnarray}
The pair of solutions (\ref{quantic NLO solution1}) is called 
the Alfven waves with the subluminal speed $0\leq v_A \leq 1$. 
The other two are damping spin modes that relax to the magnitude 
determined by the thermal vorticity. This is required by the second law of thermodynamics in such a way that the entropy production by the rotational viscosity ceases. 
The thermal vorticity, and thus spin density, is vanishing 
in the current equilibrium configuration.

The solutions for the second equation in Eq.~(\ref{eq:matrix-eqs}) are found to be 
\begin{subequations}
\label{quartic NLO solution}
\begin{eqnarray}
\label{quartic NLO solution1}
    \omega \= \pm v_1 k+i k^2 \frac{W_1-W_2 v_1^2}{2(v_1^2-v_2^2)},  
    \\ 
    \label{quartic NLO solution2}
    \omega \= \pm v_2 k-i k^2\frac{W_1-W_2 v_2^2}{2(v_1^2-v_2^2)}, 
    \\
    \label{quartic NLO solution3}
    \omega \= -i\Gamma_\parallel-{ ik^2 \left(  {W_3}-  {W_2}\right)}, 
\end{eqnarray}
\end{subequations}
with the explicit forms of the phase velocities and of $W_i$:  
\begin{subequations}
\label{eq:W-sol}
\begin{eqnarray} 
 v_{1,2} \=\frac{1}{\sqrt{2}}\Big[ \, v_A^2+ c_s^2(1-v_A^2\sin^2 \theta
    \pm\sqrt{\big (v_A^2+ c_s^2(1-v_A^2\sin^2 \theta)\big)^2 -4v_A^2 c_s^2 \cos^2\theta} \, \Big]^{\frac 12},
    \nnb
    \label{v_12}
 \\
W_1  \= \frac{1}{h}\Big[
\tilde \eta_\para
\Big( c_s^2  \cos^2 (2\theta)
+  \frac{v_A^2  \sin^2 \theta}{1-v_A^2}
 \Big)  
+ h \rho'_\perp c_s^2 \frac{ 1 - v_A^2 \cos^2 \theta }{1-v_A^2}  + 
\zeta_\para   \frac{v_A^2}{1-v_A^2}  \cos^2 \theta
\nnb
&& 
+ ( \zeta_\para +\zeta_\perp- 2 \zeta_\times 
+ \eta_\perp ) c_s^2  \sin^2 \theta \cos^2 \theta \Big]
,  
\\
W_2 \=  \frac{1}{h}\Big[
\tilde \eta_\para
\frac{1 - v_A^2  \cos^2 \theta}{1-v_A^2}
+  \zeta_\para\frac{\cos^2 \theta }{1-v_A^2}  
+ ( \zeta_\perp  + \eta_\perp ) \sin^2\theta
+ h\rho'_\perp \Big( 1 + \frac{v_A^2  c_s^2 }{1-v_A^2} \sin^2\theta  \Big)
\Big]
,
\nnb\\
W_3 \= \frac{1}{h}\Big[\cos^2\theta(\eta_\parallel+\gamma_\parallel-2\xi_\parallel) 
+\frac{\sin^2\theta}{1-v_A^2}(\eta_\parallel+\gamma_\parallel+2\xi_\parallel)
+\zeta_\parallel\frac{\cos^2\theta}{1-v_A^2}
\nnb
&&
+(\zeta_\perp+\eta_\perp)\sin^2\theta 
+ h\rho_\perp^\prime(1+\frac{v_A^2 c_s^2}{1-v_A^2}\sin^2\theta)\Big]
.
\end{eqnarray}\label{quartic NLO W}
\end{subequations}
We find two pairs of propagating modes in Eqs.~(\ref{quartic NLO solution1}) and (\ref{quartic NLO solution2}) which are known as the fast and slow magneto-sonic waves, respectively. 
We showed that $0\leq v_{1,2} \leq 1$ when $0 \leq c_s \leq 1$ and  $0\leq v_A \leq 1$ \cite{Fang:2024skm}.

We also showed that all of the nine modes are damping in time 
as long as the thermodynamic inequalities (\ref{inequality}) are satisfied 
\cite{Fang:2024skm, Fang:2024sym}. 
Note, however, that this can be an observer-dependent statement when causality is not guaranteed due to dissipative effects 
(see discussions in Ref.~\cite{Fang:2024skm} and references therein).

 \begin{figure}[t!]
\begin{minipage}{0.5\hsize} 
	\begin{center} 
\includegraphics[width=\hsize]{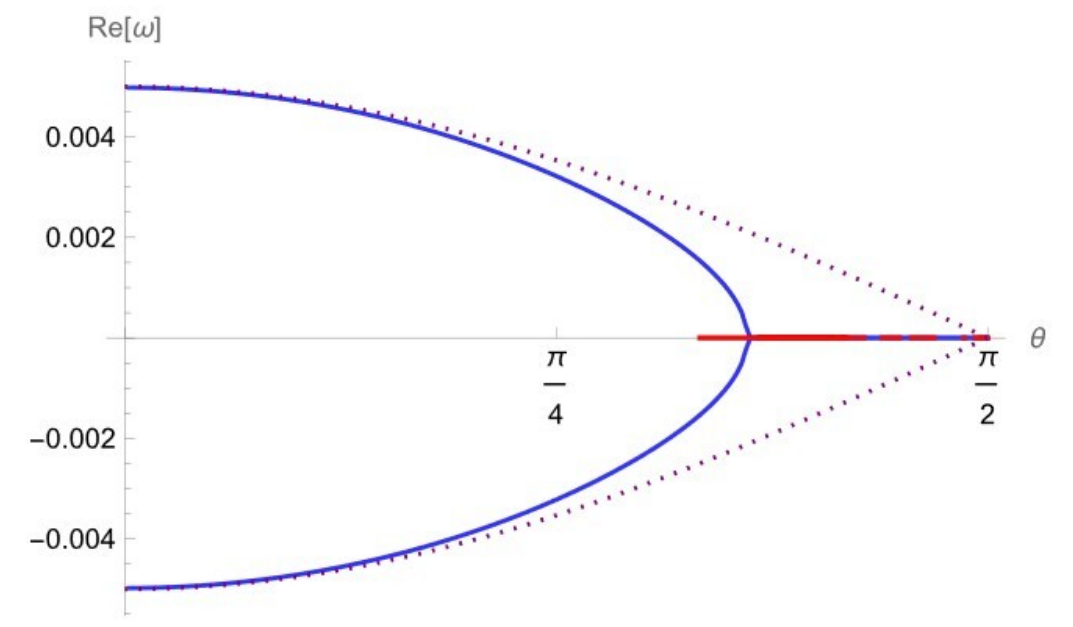}
	\end{center}
\end{minipage}
\begin{minipage}{0.5\hsize}
	\begin{center}
\includegraphics[width=\hsize]{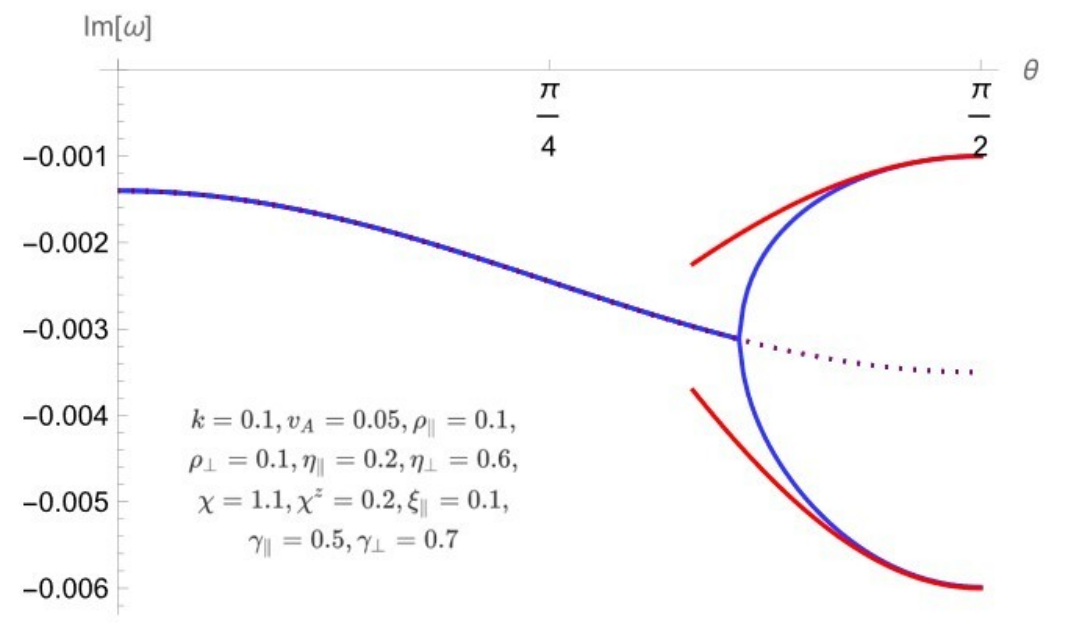}
	\end{center}
\end{minipage}
\caption{The dispersion relations for the Alfven modes. Blue curves show the ``exact solution'' without any expansion. 
    Dotted curves show the small-$k$ expansion in Eq.~(\ref{NLO11}). 
    Red curves show the small-cosine expansion. 
 }
\label{fig:spin-MHD-4}
\begin{minipage}{0.5\hsize} 
	\begin{center} 
\includegraphics[width=\hsize]{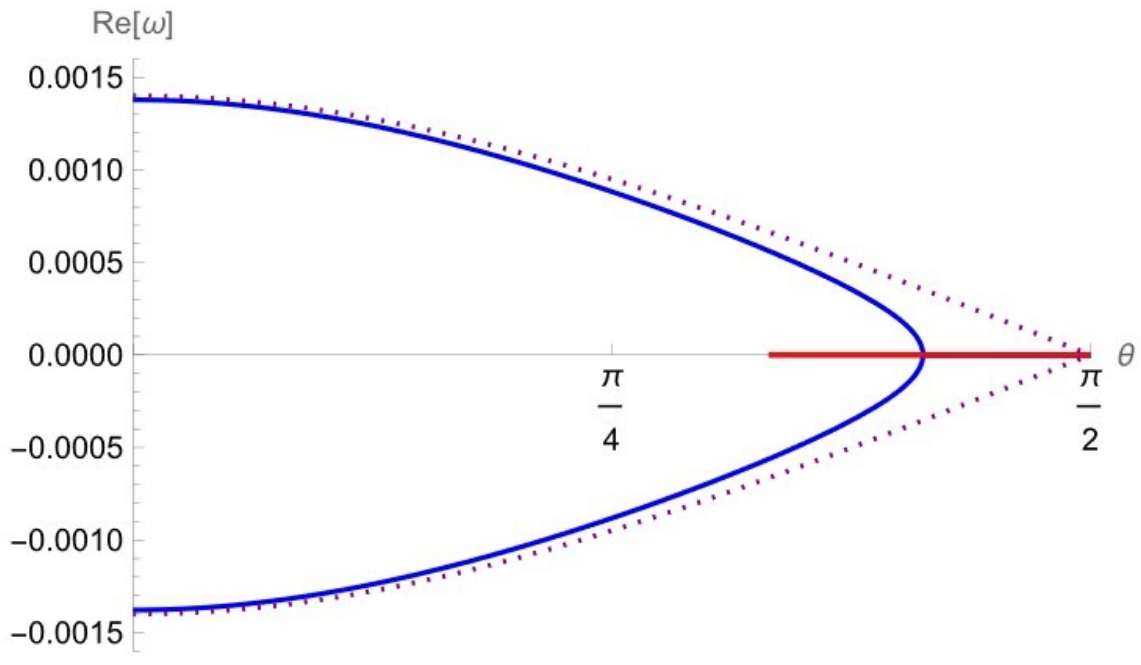}
	\end{center}
\end{minipage}
\begin{minipage}{0.5\hsize}
	\begin{center}
\includegraphics[width=\hsize]{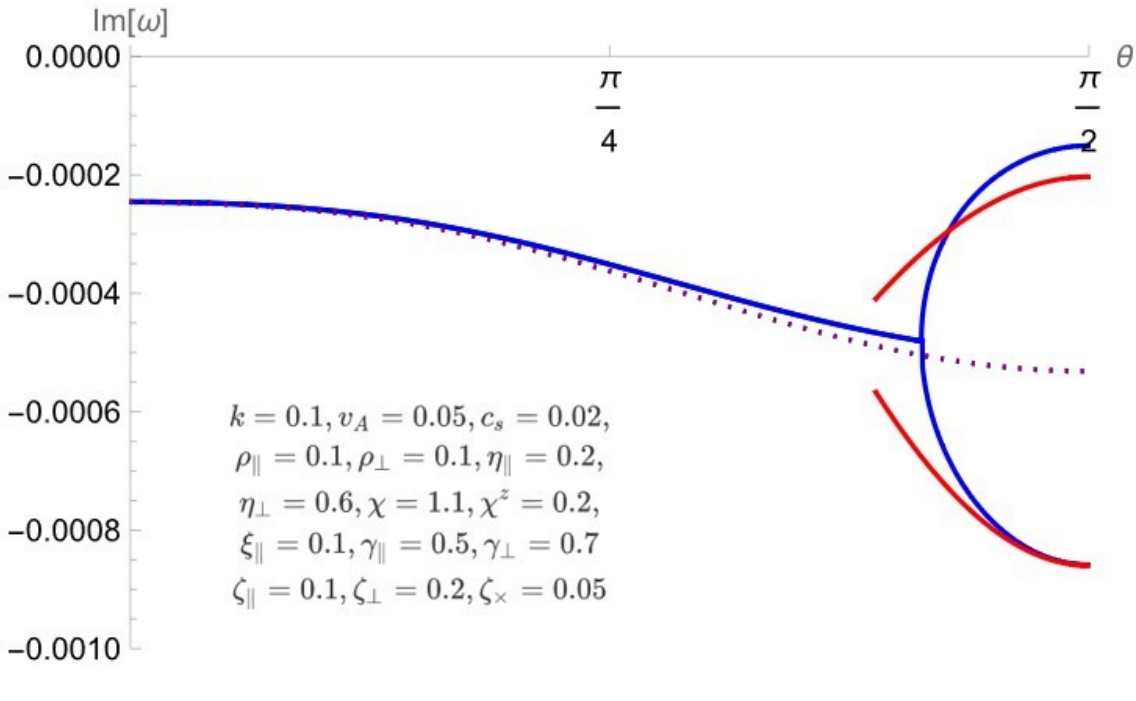}
	\end{center}
\end{minipage}
\caption{The dispersion relations for the slow magneto-sonic modes. The blue curves represent the numerical solutions of the quartic equation. 
   Dotted curves show the small-$k$ expansion in Eq.~(\ref{quartic NLO solution2}). 
    Red curves show the small-cosine expansion. 
 }
\label{fig:spin-MHD-5}
\end{figure}

\subsection{Breakdown of the small-momentum expansion for MHD}

We have obtained the solutions for an arbitrary angle $\theta$ up to the $k^2$ order in the above. 
However, we point out that the small-$k$ expansion breaks down depending on the angle $\theta$. 
One can easily realize the breakdown of the small-$k$ expansion by comparing the solutions in Eqs.~(\ref{quantic NLO solution}) and (\ref{quartic NLO solution}) with the solutions at $\theta = \pi/2$; 
The latter solutions can be obtained without any expansion. 
The small-$k$ expansion breaks down because the $n$-th order term blows up as $\sim 1/\cos^n \theta$ when $ \cos\theta \to 0$ (see below and Ref.~\cite{Fang:2024skm} for details).

We show the angle dependence of the solutions with the numerical plots in Fig.~\ref{fig:spin-MHD-4} for the Alfven waves 
and Fig.~\ref{fig:spin-MHD-5} for the slow magneto-sonic waves. 
Numerical solutions indicate the presence of a {\it critical angle} where 
a pair of propagating modes turns into two distinct diffusive modes. 
The small-$ k$ expansion shown with dotted lines significantly deviates 
from the numerical solutions shown in blue 
as $ \theta$ approaches the critical angle. 
Beyond the critical angle, the small-$ k$ expansion does not work at all.

This singular behavior was recently recognized in MHD 
without spin \cite{Fang:2024skm}. 
This behavior can be confirmed with the solutions for the Alfven waves 
that can be obtained without any expansion as 
\begin{eqnarray}
\label{eq:sol-Alfven}
\omega  \= \pm \sqrt{v_A^2  k^2 \cos^2  \theta
- \frac14  (\tilde \rho - \tilde \eta )^2k^4  }
- \frac{i}{2} (\tilde \rho + \tilde \eta) k^2
, 
\end{eqnarray}
where $\tilde \rho =  \rho'_\perp \cos^2 \theta
+\rho'_\para  \sin^2 \theta$ and $
\tilde \eta =\frac{1}{h}( \eta_\para \cos^2 \theta + \eta_\perp \sin^2 \theta ) $. 
The square root in Eq.~(\ref{eq:sol-Alfven}) becomes a pure imaginary number when $k$ grows larger than $2v_A\left|\frac{\cos\theta}{\tilde \rho-\tilde\eta}\right|$. 
Namely, the dispersion relation loses the real part, turning to purely diffusive modes. 
Clearly, this behavior is different from that of the solution in the small-$k$ expansion (\ref{quantic NLO solution1}) that always has the nonzero Alfven velocity. 
This is the breakdown of the momentum expansion beyond the critical angle seen in Fig.~\ref{fig:spin-MHD-4}. 
One finds the same issue in the slow magneto-sonic waves as seen in Fig.~\ref{fig:spin-MHD-5}. Spin MHD shares the same issue originating from MHD. 
However, damping spin modes do not induce additional issues.

Beyond the critical angle, we organize another series 
in the order of $\cos\theta $. 
These results are shown in red in the figures. 
They reproduce the numerical results when $\cos\theta $ is small. 
The deviation seen in Fig.~\ref{fig:spin-MHD-5} near $ \theta=\pi/2$ is 
due to the $ k$ expansion performed after the cosine expansion.\footnote{ 
One gets a better agreement if one includes the higher $ k$ terms, 
which however does not mean a more accurate solution since the higher $ k$ terms are uncertain in the first-order hydrodynamics. 
}
The reader is referred to Refs.~\cite{Fang:2024skm,Fang:2024sym} for explicit forms of the cosine expansions.

\section{conclusion}\label{s5}

We derived the constitutive equations with a broken spatial rotational 
symmetry due to the presence of a magnetic field. 
This spatial anisotroy induces the two distinct components of 
shear and bulk viscous coefficients and of resistivity. 
Also, an additional viscous coefficient appears as the cross bulk viscosity. 
Including spin degrees of freedom, we find the two distinct components of the rotational viscous coefficients and the cross term that converts a vorticity into the symmetric stress and a shear into the antisymmetric torque.

We elaborated on the angle dependence of the linear waves 
with a complete set of analytic solutions. 
These solutions indicate that the small-momentum expansion is spoiled due to the blow-up of the higher-order terms. 
Numerical solutions indicate that there is a critical angle where the two competing expansion parameters, the momentum $k$ and $\cos \theta$, have similar magnitudes. 
This non-analytic structure is potentially 
a general issue in anisotropic systems.
Spin modes are damping in time in the absence of an order-one thermal vorticity. 
The stability of these modes within the fluid rest frame 
is discussed in Ref.~\cite{Fang:2024skm, Fang:2024sym}.

\section*{Acknowledgements}
This contribution is partially supported by the JSPS KAKENHI under grant Nos. 20K03948 and 22H01216.  



\bibliographystyle{elsarticle-num-names}
\bibliography{ref}

\begin{thebibliography}{48}
\expandafter\ifx\csname natexlab\endcsname\relax\def\natexlab#1{#1}\fi
\providecommand{\url}[1]{\texttt{#1}}
\providecommand{\href}[2]{#2}
\providecommand{\path}[1]{#1}
\providecommand{\DOIprefix}{doi:}
\providecommand{\ArXivprefix}{arXiv:}
\providecommand{\URLprefix}{URL: }
\providecommand{\Pubmedprefix}{pmid:}
\providecommand{\doi}[1]{\href{http://dx.doi.org/#1}{\path{#1}}}
\providecommand{\Pubmed}[1]{\href{pmid:#1}{\path{#1}}}
\providecommand{\bibinfo}[2]{#2}
\ifx\xfnm\relax \def\xfnm[#1]{\unskip,\space#1}\fi
\bibitem[{Fang et~al.(2024{\natexlab{a}})Fang, Hattori, and Hu}]{Fang:2024skm}
\bibinfo{author}{Z.~Fang}, \bibinfo{author}{K.~Hattori},
  \bibinfo{author}{J.~Hu},
\newblock \bibinfo{title}{{Analytic solutions for the linearized first-order
  magnetohydrodynamics and implications for causality and stability}},
\newblock \bibinfo{journal}{Phys. Rev. D} \bibinfo{volume}{110}
  (\bibinfo{year}{2024}{\natexlab{a}}) \bibinfo{pages}{056049}.
  \DOIprefix\doi{10.1103/PhysRevD.110.056049}.
  \href{http://arxiv.org/abs/2402.18601}{{\tt arXiv:2402.18601}}.
\bibitem[{Fang et~al.(2024{\natexlab{b}})Fang, Hattori, and Hu}]{Fang:2024sym}
\bibinfo{author}{Z.~Fang}, \bibinfo{author}{K.~Hattori},
  \bibinfo{author}{J.~Hu},
\newblock \bibinfo{title}{{Anisotropic linear waves and breakdown of the
  momentum expansion in spin magnetohydrodynamics}}
  (\bibinfo{year}{2024}{\natexlab{b}}).
  \href{http://arxiv.org/abs/2409.07096}{{\tt arXiv:2409.07096}}.
\bibitem[{Kharzeev et~al.(2016)Kharzeev, Liao, Voloshin, and
  Wang}]{Kharzeev:2015znc}
\bibinfo{author}{D.~E. Kharzeev}, \bibinfo{author}{J.~Liao},
  \bibinfo{author}{S.~A. Voloshin}, \bibinfo{author}{G.~Wang},
\newblock \bibinfo{title}{{Chiral magnetic and vortical effects in high-energy
  nuclear collisions: A status report}},
\newblock \bibinfo{journal}{Prog. Part. Nucl. Phys.} \bibinfo{volume}{88}
  (\bibinfo{year}{2016}) \bibinfo{pages}{1--28}.
  \DOIprefix\doi{10.1016/j.ppnp.2016.01.001}.
  \href{http://arxiv.org/abs/1511.04050}{{\tt arXiv:1511.04050}}.
\bibitem[{Hattori and Huang(2017)}]{Hattori:2016emy}
\bibinfo{author}{K.~Hattori}, \bibinfo{author}{X.-G. Huang},
\newblock \bibinfo{title}{{Novel quantum phenomena induced by strong magnetic
  fields in heavy-ion collisions}},
\newblock \bibinfo{journal}{Nucl. Sci. Tech.} \bibinfo{volume}{28}
  (\bibinfo{year}{2017}) \bibinfo{pages}{26}.
  \DOIprefix\doi{10.1007/s41365-016-0178-3}.
  \href{http://arxiv.org/abs/1609.00747}{{\tt arXiv:1609.00747}}.
\bibitem[{Yan and Felser(2017)}]{Yan:2016euz}
\bibinfo{author}{B.~Yan}, \bibinfo{author}{C.~Felser},
\newblock \bibinfo{title}{{Topological Materials: Weyl Semimetals}},
\newblock \bibinfo{journal}{Ann. Rev. Condensed Matter Phys.}
  \bibinfo{volume}{8} (\bibinfo{year}{2017}) \bibinfo{pages}{337--354}.
  \DOIprefix\doi{10.1146/annurev-conmatphys-031016-025458}.
  \href{http://arxiv.org/abs/1611.04182}{{\tt arXiv:1611.04182}}.
\bibitem[{Armitage et~al.(2018)Armitage, Mele, and
  Vishwanath}]{Armitage:2017cjs}
\bibinfo{author}{N.~P. Armitage}, \bibinfo{author}{E.~J. Mele},
  \bibinfo{author}{A.~Vishwanath},
\newblock \bibinfo{title}{{Weyl and Dirac Semimetals in Three Dimensional
  Solids}},
\newblock \bibinfo{journal}{Rev. Mod. Phys.} \bibinfo{volume}{90}
  (\bibinfo{year}{2018}) \bibinfo{pages}{015001}.
  \DOIprefix\doi{10.1103/RevModPhys.90.015001}.
  \href{http://arxiv.org/abs/1705.01111}{{\tt arXiv:1705.01111}}.
\bibitem[{Kamada et~al.(2023)Kamada, Yamamoto, and Yang}]{Kamada:2022nyt}
\bibinfo{author}{K.~Kamada}, \bibinfo{author}{N.~Yamamoto},
  \bibinfo{author}{D.-L. Yang},
\newblock \bibinfo{title}{{Chiral effects in astrophysics and cosmology}},
\newblock \bibinfo{journal}{Prog. Part. Nucl. Phys.} \bibinfo{volume}{129}
  (\bibinfo{year}{2023}) \bibinfo{pages}{104016}.
  \DOIprefix\doi{10.1016/j.ppnp.2022.104016}.
  \href{http://arxiv.org/abs/2207.09184}{{\tt arXiv:2207.09184}}.
\bibitem[{Son and Surowka(2009)}]{Son:2009tf}
\bibinfo{author}{D.~T. Son}, \bibinfo{author}{P.~Surowka},
\newblock \bibinfo{title}{{Hydrodynamics with Triangle Anomalies}},
\newblock \bibinfo{journal}{Phys. Rev. Lett.} \bibinfo{volume}{103}
  (\bibinfo{year}{2009}) \bibinfo{pages}{191601}.
  \DOIprefix\doi{10.1103/PhysRevLett.103.191601}.
  \href{http://arxiv.org/abs/0906.5044}{{\tt arXiv:0906.5044}}.
\bibitem[{Grozdanov et~al.(2017)Grozdanov, Hofman, and
  Iqbal}]{Grozdanov:2016tdf}
\bibinfo{author}{S.~Grozdanov}, \bibinfo{author}{D.~M. Hofman},
  \bibinfo{author}{N.~Iqbal},
\newblock \bibinfo{title}{{Generalized global symmetries and dissipative
  magnetohydrodynamics}},
\newblock \bibinfo{journal}{Phys. Rev. D} \bibinfo{volume}{95}
  (\bibinfo{year}{2017}) \bibinfo{pages}{096003}.
  \DOIprefix\doi{10.1103/PhysRevD.95.096003}.
  \href{http://arxiv.org/abs/1610.07392}{{\tt arXiv:1610.07392}}.
\bibitem[{Hattori et~al.(2019)Hattori, Hirono, Yee, and Yin}]{Hattori:2017usa}
\bibinfo{author}{K.~Hattori}, \bibinfo{author}{Y.~Hirono},
  \bibinfo{author}{H.-U. Yee}, \bibinfo{author}{Y.~Yin},
\newblock \bibinfo{title}{{MagnetoHydrodynamics with chiral anomaly: phases of
  collective excitations and instabilities}},
\newblock \bibinfo{journal}{Phys. Rev. D} \bibinfo{volume}{100}
  (\bibinfo{year}{2019}) \bibinfo{pages}{065023}.
  \DOIprefix\doi{10.1103/PhysRevD.100.065023}.
  \href{http://arxiv.org/abs/1711.08450}{{\tt arXiv:1711.08450}}.
\bibitem[{Hongo et~al.(2021)Hongo, Huang, Kaminski, Stephanov, and
  Yee}]{Hongo:2021ona}
\bibinfo{author}{M.~Hongo}, \bibinfo{author}{X.-G. Huang},
  \bibinfo{author}{M.~Kaminski}, \bibinfo{author}{M.~Stephanov},
  \bibinfo{author}{H.-U. Yee},
\newblock \bibinfo{title}{{Relativistic spin hydrodynamics with torsion and
  linear response theory for spin relaxation}},
\newblock \bibinfo{journal}{JHEP} \bibinfo{volume}{11} (\bibinfo{year}{2021})
  \bibinfo{pages}{150}. \DOIprefix\doi{10.1007/JHEP11(2021)150}.
  \href{http://arxiv.org/abs/2107.14231}{{\tt arXiv:2107.14231}}.
\bibitem[{Matsumoto et~al.(2022)Matsumoto, Yamamoto, and
  Yang}]{Matsumoto:2022lyb}
\bibinfo{author}{J.~Matsumoto}, \bibinfo{author}{N.~Yamamoto},
  \bibinfo{author}{D.-L. Yang},
\newblock \bibinfo{title}{{Chiral plasma instability and inverse cascade from
  nonequilibrium left-handed neutrinos in core-collapse supernovae}},
\newblock \bibinfo{journal}{Phys. Rev. D} \bibinfo{volume}{105}
  (\bibinfo{year}{2022}) \bibinfo{pages}{123029}.
  \DOIprefix\doi{10.1103/PhysRevD.105.123029}.
  \href{http://arxiv.org/abs/2202.09205}{{\tt arXiv:2202.09205}}.
\bibitem[{Brandenburg et~al.(2023)Brandenburg, Kamada, Mukaida, Schmitz, and
  Schober}]{Brandenburg:2023aco}
\bibinfo{author}{A.~Brandenburg}, \bibinfo{author}{K.~Kamada},
  \bibinfo{author}{K.~Mukaida}, \bibinfo{author}{K.~Schmitz},
  \bibinfo{author}{J.~Schober},
\newblock \bibinfo{title}{{Chiral magnetohydrodynamics with zero total
  chirality}},
\newblock \bibinfo{journal}{Phys. Rev. D} \bibinfo{volume}{108}
  (\bibinfo{year}{2023}) \bibinfo{pages}{063529}.
  \DOIprefix\doi{10.1103/PhysRevD.108.063529}.
  \href{http://arxiv.org/abs/2304.06612}{{\tt arXiv:2304.06612}}.
\bibitem[{Hattori et~al.(2022)Hattori, Hongo, and Huang}]{Hattori:2022hyo}
\bibinfo{author}{K.~Hattori}, \bibinfo{author}{M.~Hongo},
  \bibinfo{author}{X.-G. Huang},
\newblock \bibinfo{title}{{New Developments in Relativistic
  Magnetohydrodynamics}},
\newblock \bibinfo{journal}{Symmetry} \bibinfo{volume}{14}
  (\bibinfo{year}{2022}) \bibinfo{pages}{1851}.
  \DOIprefix\doi{10.3390/sym14091851}.
  \href{http://arxiv.org/abs/2207.12794}{{\tt arXiv:2207.12794}}.
\bibitem[{Hattori et~al.(2023)Hattori, Itakura, and Ozaki}]{Hattori:2023egw}
\bibinfo{author}{K.~Hattori}, \bibinfo{author}{K.~Itakura},
  \bibinfo{author}{S.~Ozaki},
\newblock \bibinfo{title}{{Strong-field physics in QED and QCD: From
  fundamentals to applications}},
\newblock \bibinfo{journal}{Prog. Part. Nucl. Phys.} \bibinfo{volume}{133}
  (\bibinfo{year}{2023}) \bibinfo{pages}{104068}.
  \DOIprefix\doi{10.1016/j.ppnp.2023.104068}.
  \href{http://arxiv.org/abs/2305.03865}{{\tt arXiv:2305.03865}}.
\bibitem[{Adamczyk et~al.(2017)}]{STAR:2017ckg}
\bibinfo{author}{L.~Adamczyk}, et~al. (\bibinfo{collaboration}{STAR}),
\newblock \bibinfo{title}{{Global $\Lambda$ hyperon polarization in nuclear
  collisions: evidence for the most vortical fluid}},
\newblock \bibinfo{journal}{Nature} \bibinfo{volume}{548}
  (\bibinfo{year}{2017}) \bibinfo{pages}{62--65}.
  \DOIprefix\doi{10.1038/nature23004}.
  \href{http://arxiv.org/abs/1701.06657}{{\tt arXiv:1701.06657}}.
\bibitem[{Adam et~al.(2018)}]{STAR:2018gyt}
\bibinfo{author}{J.~Adam}, et~al. (\bibinfo{collaboration}{STAR}),
\newblock \bibinfo{title}{{Global polarization of $\Lambda$ hyperons in Au+Au
  collisions at $\sqrt{s_{_{NN}}}$ = 200 GeV}},
\newblock \bibinfo{journal}{Phys. Rev. C} \bibinfo{volume}{98}
  (\bibinfo{year}{2018}) \bibinfo{pages}{014910}.
  \DOIprefix\doi{10.1103/PhysRevC.98.014910}.
  \href{http://arxiv.org/abs/1805.04400}{{\tt arXiv:1805.04400}}.
\bibitem[{Mohanty et~al.(2021)Mohanty, Kundu, Singha, and
  Singh}]{mohanty_spin_2021}
\bibinfo{author}{B.~Mohanty}, \bibinfo{author}{S.~Kundu},
  \bibinfo{author}{S.~Singha}, \bibinfo{author}{R.~Singh},
\newblock \bibinfo{title}{{Spin alignment measurement of vector mesons produced
  in high energy collisions}},
\newblock \bibinfo{journal}{Mod. Phys. Lett. A} \bibinfo{volume}{36}
  (\bibinfo{year}{2021}) \bibinfo{pages}{2130026}.
  \DOIprefix\doi{10.1142/S0217732321300263}.
  \href{http://arxiv.org/abs/2112.04816}{{\tt arXiv:2112.04816}}.
\bibitem[{Adam et~al.(2019)}]{STAR:2019erd}
\bibinfo{author}{J.~Adam}, et~al. (\bibinfo{collaboration}{STAR}),
\newblock \bibinfo{title}{{Polarization of $\Lambda$ ($\bar{\Lambda}$) hyperons
  along the beam direction in Au+Au collisions at $\sqrt{s_{_{NN}}}$ = 200
  GeV}},
\newblock \bibinfo{journal}{Phys. Rev. Lett.} \bibinfo{volume}{123}
  (\bibinfo{year}{2019}) \bibinfo{pages}{132301}.
  \DOIprefix\doi{10.1103/PhysRevLett.123.132301}.
  \href{http://arxiv.org/abs/1905.11917}{{\tt arXiv:1905.11917}}.
\bibitem[{Adam et~al.(2021)}]{STAR:2020xbm}
\bibinfo{author}{J.~Adam}, et~al. (\bibinfo{collaboration}{STAR}),
\newblock \bibinfo{title}{{Global Polarization of $\Xi$ and $\Omega$ Hyperons
  in Au+Au Collisions at $\sqrt {s_{NN}}$ = 200 GeV}},
\newblock \bibinfo{journal}{Phys. Rev. Lett.} \bibinfo{volume}{126}
  (\bibinfo{year}{2021}) \bibinfo{pages}{162301}.
  \DOIprefix\doi{10.1103/PhysRevLett.126.162301}.
  \href{http://arxiv.org/abs/2012.13601}{{\tt arXiv:2012.13601}}.
\bibitem[{Acharya et~al.(2023)}]{ALICE:2022dyy}
\bibinfo{author}{S.~Acharya}, et~al. (\bibinfo{collaboration}{ALICE}),
\newblock \bibinfo{title}{{Measurement of the J/\ensuremath{\psi} Polarization
  with Respect to the Event Plane in Pb-Pb Collisions at the LHC}},
\newblock \bibinfo{journal}{Phys. Rev. Lett.} \bibinfo{volume}{131}
  (\bibinfo{year}{2023}) \bibinfo{pages}{042303}.
  \DOIprefix\doi{10.1103/PhysRevLett.131.042303}.
  \href{http://arxiv.org/abs/2204.10171}{{\tt arXiv:2204.10171}}.
\bibitem[{Micheletti(2023)}]{Micheletti:2023qlh}
\bibinfo{author}{L.~Micheletti} (\bibinfo{collaboration}{ALICE}),
\newblock \bibinfo{title}{{Vector Mesons Polarization in Pb\textendash{}Pb and
  $pp$ Collisions with ALICE}},
\newblock \bibinfo{journal}{Acta Phys. Polon. Supp.} \bibinfo{volume}{16}
  (\bibinfo{year}{2023}) \bibinfo{pages}{1--A35}.
  \DOIprefix\doi{10.5506/APhysPolBSupp.16.1-A35}.
\bibitem[{Liang and Wang(2005{\natexlab{a}})}]{Liang:2004ph}
\bibinfo{author}{Z.-T. Liang}, \bibinfo{author}{X.-N. Wang},
\newblock \bibinfo{title}{{Globally polarized quark-gluon plasma in non-central
  A+A collisions}},
\newblock \bibinfo{journal}{Phys. Rev. Lett.} \bibinfo{volume}{94}
  (\bibinfo{year}{2005}{\natexlab{a}}) \bibinfo{pages}{102301}.
  \DOIprefix\doi{10.1103/PhysRevLett.94.102301}.
  \href{http://arxiv.org/abs/nucl-th/0410079}{{\tt arXiv:nucl-th/0410079}},
  \bibinfo{note}{[Erratum: Phys.Rev.Lett. 96, 039901 (2006)]}.
\bibitem[{Liang and Wang(2005{\natexlab{b}})}]{Liang:2004xn}
\bibinfo{author}{Z.-T. Liang}, \bibinfo{author}{X.-N. Wang},
\newblock \bibinfo{title}{{Spin alignment of vector mesons in non-central A+A
  collisions}},
\newblock \bibinfo{journal}{Phys. Lett. B} \bibinfo{volume}{629}
  (\bibinfo{year}{2005}{\natexlab{b}}) \bibinfo{pages}{20--26}.
  \DOIprefix\doi{10.1016/j.physletb.2005.09.060}.
  \href{http://arxiv.org/abs/nucl-th/0411101}{{\tt arXiv:nucl-th/0411101}}.
\bibitem[{Betz et~al.(2007)Betz, Gyulassy, and Torrieri}]{Betz:2007kg}
\bibinfo{author}{B.~Betz}, \bibinfo{author}{M.~Gyulassy},
  \bibinfo{author}{G.~Torrieri},
\newblock \bibinfo{title}{{Polarization probes of vorticity in heavy ion
  collisions}},
\newblock \bibinfo{journal}{Phys. Rev. C} \bibinfo{volume}{76}
  (\bibinfo{year}{2007}) \bibinfo{pages}{044901}.
  \DOIprefix\doi{10.1103/PhysRevC.76.044901}.
  \href{http://arxiv.org/abs/0708.0035}{{\tt arXiv:0708.0035}}.
\bibitem[{Becattini et~al.(2008)Becattini, Piccinini, and
  Rizzo}]{Becattini:2007sr}
\bibinfo{author}{F.~Becattini}, \bibinfo{author}{F.~Piccinini},
  \bibinfo{author}{J.~Rizzo},
\newblock \bibinfo{title}{{Angular momentum conservation in heavy ion
  collisions at very high energy}},
\newblock \bibinfo{journal}{Phys. Rev. C} \bibinfo{volume}{77}
  (\bibinfo{year}{2008}) \bibinfo{pages}{024906}.
  \DOIprefix\doi{10.1103/PhysRevC.77.024906}.
  \href{http://arxiv.org/abs/0711.1253}{{\tt arXiv:0711.1253}}.
\bibitem[{Gao et~al.(2008)Gao, Chen, Deng, Liang, Wang, and Wang}]{Gao:2007bc}
\bibinfo{author}{J.-H. Gao}, \bibinfo{author}{S.-W. Chen},
  \bibinfo{author}{W.-t. Deng}, \bibinfo{author}{Z.-T. Liang},
  \bibinfo{author}{Q.~Wang}, \bibinfo{author}{X.-N. Wang},
\newblock \bibinfo{title}{{Global quark polarization in non-central A+A
  collisions}},
\newblock \bibinfo{journal}{Phys. Rev. C} \bibinfo{volume}{77}
  (\bibinfo{year}{2008}) \bibinfo{pages}{044902}.
  \DOIprefix\doi{10.1103/PhysRevC.77.044902}.
  \href{http://arxiv.org/abs/0710.2943}{{\tt arXiv:0710.2943}}.
\bibitem[{Becattini et~al.(2017)Becattini, Karpenko, Lisa, Upsal, and
  Voloshin}]{Becattini:2016gvu}
\bibinfo{author}{F.~Becattini}, \bibinfo{author}{I.~Karpenko},
  \bibinfo{author}{M.~Lisa}, \bibinfo{author}{I.~Upsal},
  \bibinfo{author}{S.~Voloshin},
\newblock \bibinfo{title}{{Global hyperon polarization at local thermodynamic
  equilibrium with vorticity, magnetic field and feed-down}},
\newblock \bibinfo{journal}{Phys. Rev. C} \bibinfo{volume}{95}
  (\bibinfo{year}{2017}) \bibinfo{pages}{054902}.
  \DOIprefix\doi{10.1103/PhysRevC.95.054902}.
  \href{http://arxiv.org/abs/1610.02506}{{\tt arXiv:1610.02506}}.
\bibitem[{Becattini et~al.(2021)Becattini, Buzzegoli, Inghirami, Karpenko, and
  Palermo}]{Becattini:2021iol}
\bibinfo{author}{F.~Becattini}, \bibinfo{author}{M.~Buzzegoli},
  \bibinfo{author}{G.~Inghirami}, \bibinfo{author}{I.~Karpenko},
  \bibinfo{author}{A.~Palermo},
\newblock \bibinfo{title}{{Local Polarization and Isothermal Local Equilibrium
  in Relativistic Heavy Ion Collisions}},
\newblock \bibinfo{journal}{Phys. Rev. Lett.} \bibinfo{volume}{127}
  (\bibinfo{year}{2021}) \bibinfo{pages}{272302}.
  \DOIprefix\doi{10.1103/PhysRevLett.127.272302}.
  \href{http://arxiv.org/abs/2103.14621}{{\tt arXiv:2103.14621}}.
\bibitem[{Wu et~al.(2019)Wu, Pang, Huang, and Wang}]{Wu:2019eyi}
\bibinfo{author}{H.-Z. Wu}, \bibinfo{author}{L.-G. Pang},
  \bibinfo{author}{X.-G. Huang}, \bibinfo{author}{Q.~Wang},
\newblock \bibinfo{title}{{Local spin polarization in high energy heavy ion
  collisions}},
\newblock \bibinfo{journal}{Phys. Rev. Research.} \bibinfo{volume}{1}
  (\bibinfo{year}{2019}) \bibinfo{pages}{033058}.
  \DOIprefix\doi{10.1103/PhysRevResearch.1.033058}.
  \href{http://arxiv.org/abs/1906.09385}{{\tt arXiv:1906.09385}}.
\bibitem[{Becattini et~al.(2024)Becattini, Buzzegoli, Niida, Pu, Tang, and
  Wang}]{Becattini:2024uha}
\bibinfo{author}{F.~Becattini}, \bibinfo{author}{M.~Buzzegoli},
  \bibinfo{author}{T.~Niida}, \bibinfo{author}{S.~Pu}, \bibinfo{author}{A.-H.
  Tang}, \bibinfo{author}{Q.~Wang},
\newblock \bibinfo{title}{{Spin polarization in relativistic heavy-ion
  collisions}}  (\bibinfo{year}{2024}).
  \href{http://arxiv.org/abs/2402.04540}{{\tt arXiv:2402.04540}}.
\bibitem[{Hattori et~al.(2019)Hattori, Hongo, Huang, Matsuo, and
  Taya}]{Hattori:2019lfp}
\bibinfo{author}{K.~Hattori}, \bibinfo{author}{M.~Hongo},
  \bibinfo{author}{X.-G. Huang}, \bibinfo{author}{M.~Matsuo},
  \bibinfo{author}{H.~Taya},
\newblock \bibinfo{title}{{Fate of spin polarization in a relativistic fluid:
  An entropy-current analysis}},
\newblock \bibinfo{journal}{Phys. Lett. B} \bibinfo{volume}{795}
  (\bibinfo{year}{2019}) \bibinfo{pages}{100--106}.
  \DOIprefix\doi{10.1016/j.physletb.2019.05.040}.
  \href{http://arxiv.org/abs/1901.06615}{{\tt arXiv:1901.06615}}.
\bibitem[{Fukushima and Pu(2021)}]{Fukushima:2020ucl}
\bibinfo{author}{K.~Fukushima}, \bibinfo{author}{S.~Pu},
\newblock \bibinfo{title}{{Spin hydrodynamics and symmetric energy-momentum
  tensors \textendash{} A current induced by the spin vorticity
  \textendash{}}},
\newblock \bibinfo{journal}{Phys. Lett. B} \bibinfo{volume}{817}
  (\bibinfo{year}{2021}) \bibinfo{pages}{136346}.
  \DOIprefix\doi{10.1016/j.physletb.2021.136346}.
  \href{http://arxiv.org/abs/2010.01608}{{\tt arXiv:2010.01608}}.
\bibitem[{Gallegos et~al.(2021)Gallegos, G\"ursoy, and
  Yarom}]{Gallegos:2021bzp}
\bibinfo{author}{A.~D. Gallegos}, \bibinfo{author}{U.~G\"ursoy},
  \bibinfo{author}{A.~Yarom},
\newblock \bibinfo{title}{{Hydrodynamics of spin currents}},
\newblock \bibinfo{journal}{SciPost Phys.} \bibinfo{volume}{11}
  (\bibinfo{year}{2021}) \bibinfo{pages}{041}.
  \DOIprefix\doi{10.21468/SciPostPhys.11.2.041}.
  \href{http://arxiv.org/abs/2101.04759}{{\tt arXiv:2101.04759}}.
\bibitem[{Li et~al.(2021)Li, Stephanov, and Yee}]{Li:2020eon}
\bibinfo{author}{S.~Li}, \bibinfo{author}{M.~A. Stephanov},
  \bibinfo{author}{H.-U. Yee},
\newblock \bibinfo{title}{{Nondissipative Second-Order Transport, Spin, and
  Pseudogauge Transformations in Hydrodynamics}},
\newblock \bibinfo{journal}{Phys. Rev. Lett.} \bibinfo{volume}{127}
  (\bibinfo{year}{2021}) \bibinfo{pages}{082302}.
  \DOIprefix\doi{10.1103/PhysRevLett.127.082302}.
  \href{http://arxiv.org/abs/2011.12318}{{\tt arXiv:2011.12318}}.
\bibitem[{Hu(2021)}]{Hu:2021lnx}
\bibinfo{author}{J.~Hu},
\newblock \bibinfo{title}{{Kubo formulae for first-order spin hydrodynamics}},
\newblock \bibinfo{journal}{Phys. Rev. D} \bibinfo{volume}{103}
  (\bibinfo{year}{2021}) \bibinfo{pages}{116015}.
  \DOIprefix\doi{10.1103/PhysRevD.103.116015}.
  \href{http://arxiv.org/abs/2101.08440}{{\tt arXiv:2101.08440}}.
\bibitem[{Hu(2023)}]{Hu:2022azy}
\bibinfo{author}{J.~Hu},
\newblock \bibinfo{title}{{Cross effects in spin hydrodynamics: Entropy
  analysis and statistical operator}},
\newblock \bibinfo{journal}{Phys. Rev. C} \bibinfo{volume}{107}
  (\bibinfo{year}{2023}) \bibinfo{pages}{024915}.
  \DOIprefix\doi{10.1103/PhysRevC.107.024915}.
  \href{http://arxiv.org/abs/2209.10979}{{\tt arXiv:2209.10979}}.
\bibitem[{Singh et~al.(2023)Singh, Shokri, and Mehr}]{Singh:2022ltu}
\bibinfo{author}{R.~Singh}, \bibinfo{author}{M.~Shokri}, \bibinfo{author}{S.~M.
  A.~T. Mehr},
\newblock \bibinfo{title}{{Relativistic hydrodynamics with spin in the presence
  of electromagnetic fields}},
\newblock \bibinfo{journal}{Nucl. Phys. A} \bibinfo{volume}{1035}
  (\bibinfo{year}{2023}) \bibinfo{pages}{122656}.
  \DOIprefix\doi{10.1016/j.nuclphysa.2023.122656}.
  \href{http://arxiv.org/abs/2202.11504}{{\tt arXiv:2202.11504}}.
\bibitem[{Cao et~al.(2022)Cao, Hattori, Hongo, Huang, and Taya}]{Cao:2022aku}
\bibinfo{author}{Z.~Cao}, \bibinfo{author}{K.~Hattori},
  \bibinfo{author}{M.~Hongo}, \bibinfo{author}{X.-G. Huang},
  \bibinfo{author}{H.~Taya},
\newblock \bibinfo{title}{{Gyrohydrodynamics: Relativistic spinful fluid with
  strong vorticity}},
\newblock \bibinfo{journal}{PTEP} \bibinfo{volume}{2022} (\bibinfo{year}{2022})
  \bibinfo{pages}{071D01}. \DOIprefix\doi{10.1093/ptep/ptac091}.
  \href{http://arxiv.org/abs/2205.08051}{{\tt arXiv:2205.08051}}.
\bibitem[{Daher et~al.(2023)Daher, Das, and Ryblewski}]{Daher:2022wzf}
\bibinfo{author}{A.~Daher}, \bibinfo{author}{A.~Das},
  \bibinfo{author}{R.~Ryblewski},
\newblock \bibinfo{title}{{Stability studies of first-order spin-hydrodynamic
  frameworks}},
\newblock \bibinfo{journal}{Phys. Rev. D} \bibinfo{volume}{107}
  (\bibinfo{year}{2023}) \bibinfo{pages}{054043}.
  \DOIprefix\doi{10.1103/PhysRevD.107.054043}.
  \href{http://arxiv.org/abs/2209.10460}{{\tt arXiv:2209.10460}}.
\bibitem[{Sarwar et~al.(2023)Sarwar, Hasanujjaman, Bhatt, Mishra, and
  Alam}]{Sarwar:2022yzs}
\bibinfo{author}{G.~Sarwar}, \bibinfo{author}{M.~Hasanujjaman},
  \bibinfo{author}{J.~R. Bhatt}, \bibinfo{author}{H.~Mishra},
  \bibinfo{author}{J.-e. Alam},
\newblock \bibinfo{title}{{Causality and stability of relativistic spin
  hydrodynamics}},
\newblock \bibinfo{journal}{Phys. Rev. D} \bibinfo{volume}{107}
  (\bibinfo{year}{2023}) \bibinfo{pages}{054031}.
  \DOIprefix\doi{10.1103/PhysRevD.107.054031}.
  \href{http://arxiv.org/abs/2209.08652}{{\tt arXiv:2209.08652}}.
\bibitem[{Kiamari et~al.(2024)Kiamari, Sadooghi, and Jafari}]{Kiamari:2023fbe}
\bibinfo{author}{M.~Kiamari}, \bibinfo{author}{N.~Sadooghi},
  \bibinfo{author}{M.~S. Jafari},
\newblock \bibinfo{title}{{Relativistic magnetohydrodynamics of a spinful and
  vortical fluid: Entropy current analysis}},
\newblock \bibinfo{journal}{Phys. Rev. D} \bibinfo{volume}{109}
  (\bibinfo{year}{2024}) \bibinfo{pages}{036024}.
  \DOIprefix\doi{10.1103/PhysRevD.109.036024}.
  \href{http://arxiv.org/abs/2310.01874}{{\tt arXiv:2310.01874}}.
\bibitem[{Xie et~al.(2023)Xie, Wang, Yang, and Pu}]{Xie:2023gbo}
\bibinfo{author}{X.-Q. Xie}, \bibinfo{author}{D.-L. Wang},
  \bibinfo{author}{C.~Yang}, \bibinfo{author}{S.~Pu},
\newblock \bibinfo{title}{{Causality and stability analysis for the minimal
  causal spin hydrodynamics}},
\newblock \bibinfo{journal}{Phys. Rev. D} \bibinfo{volume}{108}
  (\bibinfo{year}{2023}) \bibinfo{pages}{094031}.
  \DOIprefix\doi{10.1103/PhysRevD.108.094031}.
  \href{http://arxiv.org/abs/2306.13880}{{\tt arXiv:2306.13880}}.
\bibitem[{Ren et~al.(2024)Ren, Yang, Wang, and Pu}]{Ren:2024pur}
\bibinfo{author}{X.~Ren}, \bibinfo{author}{C.~Yang}, \bibinfo{author}{D.-L.
  Wang}, \bibinfo{author}{S.~Pu},
\newblock \bibinfo{title}{{Thermodynamic stability in relativistic viscous and
  spin hydrodynamics}},
\newblock \bibinfo{journal}{Phys. Rev. D} \bibinfo{volume}{110}
  (\bibinfo{year}{2024}) \bibinfo{pages}{034010}.
  \DOIprefix\doi{10.1103/PhysRevD.110.034010}.
  \href{http://arxiv.org/abs/2405.03105}{{\tt arXiv:2405.03105}}.
\bibitem[{Xu et~al.(2022)Xu, Lin, Huang, and Huang}]{Xu:2022hql}
\bibinfo{author}{K.~Xu}, \bibinfo{author}{F.~Lin}, \bibinfo{author}{A.~Huang},
  \bibinfo{author}{M.~Huang},
\newblock
  \bibinfo{title}{{\ensuremath{\Lambda}/\ensuremath{\Lambda}\textasciimacron{}
  polarization and splitting induced by rotation and magnetic field}},
\newblock \bibinfo{journal}{Phys. Rev. D} \bibinfo{volume}{106}
  (\bibinfo{year}{2022}) \bibinfo{pages}{L071502}.
  \DOIprefix\doi{10.1103/PhysRevD.106.L071502}.
  \href{http://arxiv.org/abs/2205.02420}{{\tt arXiv:2205.02420}}.
\bibitem[{Peng et~al.(2023)Peng, Wu, Wang, She, and Pu}]{Peng:2022cya}
\bibinfo{author}{H.-H. Peng}, \bibinfo{author}{S.~Wu}, \bibinfo{author}{R.-j.
  Wang}, \bibinfo{author}{D.~She}, \bibinfo{author}{S.~Pu},
\newblock \bibinfo{title}{{Anomalous magnetohydrodynamics with
  temperature-dependent electric conductivity and application to the global
  polarization}},
\newblock \bibinfo{journal}{Phys. Rev. D} \bibinfo{volume}{107}
  (\bibinfo{year}{2023}) \bibinfo{pages}{096010}.
  \DOIprefix\doi{10.1103/PhysRevD.107.096010}.
  \href{http://arxiv.org/abs/2211.11286}{{\tt arXiv:2211.11286}}.
\bibitem[{Buzzegoli(2023)}]{Buzzegoli:2022qrr}
\bibinfo{author}{M.~Buzzegoli},
\newblock \bibinfo{title}{{Spin polarization induced by magnetic field and the
  relativistic Barnett effect}},
\newblock \bibinfo{journal}{Nucl. Phys. A} \bibinfo{volume}{1036}
  (\bibinfo{year}{2023}) \bibinfo{pages}{122674}.
  \DOIprefix\doi{10.1016/j.nuclphysa.2023.122674}.
  \href{http://arxiv.org/abs/2211.04549}{{\tt arXiv:2211.04549}}.
\bibitem[{Sun and Yan(2024)}]{Sun:2024isb}
\bibinfo{author}{J.-A. Sun}, \bibinfo{author}{L.~Yan},
\newblock \bibinfo{title}{{Weak magnetic effect in quark-gluon plasma and local
  spin polarization}}  (\bibinfo{year}{2024}).
  \href{http://arxiv.org/abs/2401.07458}{{\tt arXiv:2401.07458}}.

\end{thebibliography}

\end{document}